# Multiferroic clusters: a new perspective for relaxor-type room-temperature multiferroics

*Leonard F. Henrichs\*, Oscar Cespedes, James Bennett, Joachim Landers, Soma Salamon, Christian Heuser, Tim Helbig, Oliver Gutfleisch, Doru C. Lupascu, Heiko Wende, Wolfgang Kleemann, and Andrew J. Bell*

L. F. H., Dr. J. B., Prof. A. J. B.

University of Leeds, Institute for Materials Research, Engineering Building, LS2 9JT Leeds, United Kingdom

Dr. O.C.

University of Leeds, School of Physics and Astronomy, E. C. Stoner Laboratory, LS2 9JT Leeds, United Kingdom

J. L., S.S., Dr. A. H., Prof. H.W., Prof. W.K.

University of Duisburg-Essen, Faculty of Physics and Center for Nanointegration Duisburg-Essen (CENIDE), Lotharstr. 1, 47048 Duisburg, Germany

T.H., Prof. O.G.

Technical University of Darmstadt, Materials Science and Earth Science, Alarich-Weiss-Straße 16, 64287 Darmstadt, Germany

Prof. D.C.L.

University of Duisburg-Essen, Institute for Materials Science and Center for Nanointegration Duisburg-Essen (CENIDE), Universitätsstr. 15, 45141 Essen, Germany

Keywords: multiferroics, magnetoelectric, piezoresponse force microscopy, magnetic force microscopy, multiferroic cluster

**Abstract**

Multiferroics are promising for sensor and memory applications, but despite all efforts invested in their research no single-phase material displaying both ferroelectricity and large magnetization at room-temperature has hitherto been reported. This situation has substantially been improved in the novel relaxor ferroelectric single-phase $(BiFe_{0.9}Co_{0.1}O_3)_{0.4}$-$(Bi_{1/2}K_{1/2}TiO_3)_{0.6}$, where polar nanoregions (PNR) transform into static-PNR (SPNR) as evidenced by piezoresponse force microscopy (PFM) and simultaneously enable congruent multiferroic clusters (MFC) to emerge from inherent ferrimagnetic $Bi(Fe,Co)O_3$ regions as verified by magnetic force microscopy (MFM) and secondary ion mass spectrometry (SIMS). On these MFC, exceptionally large direct and converse magnetoelectric coupling coefficients, $\alpha \approx 1.0 \times 10^{-5}$ s/m at room-temperature, were measured by PFM and MFM respectively. We expect the non-ergodic relaxor properties which are governed by the $Bi_{1/2}K_{1/2}TiO_3$ component to play a vital role in the strong ME coupling, by providing an electrically and mechanically flexible environment to MFC. This new class of non-ergodic relaxor multiferroics bears great potential for applications. Especially the prospect of a ME nanodot storage device seems appealing.

## 1. Introduction

After a climax in research on multiferroics and magnetoelectrics in the 1970s followed by a decline in the subsequent two decades, there has been a steep rise in the number of publications in this area since 2000, which is still ongoing.[1, 2] Major fields for applications are in sensors and logical devices such as magnetic sensors[3, 4], magnetoelectric (ME) memory[5] and voltage-driven magnetic tunnel-junctions[6].

However, one of the major obstacles is that most single-phase multiferroics only exhibit ME coupling well below room-temperature, which in combination with their relatively low coupling-coefficients between $10^{-11} – 10^{-10}$ s/m,[1, 7] makes them unsuitable for practical devices so far. Although recent attempts to prepare room-temperature single-phase multiferroics by choosing solid solutions of perovskites or Aurivillius phases have been quite successful,[8] all of them have still unacceptably weak ME coupling coefficients as to become candidates for technical applications.

In contrast, composite multiferroics, which consist of separated ferroelectric (FE) and ferromagnetic (FM) phases, exhibit much higher coupling-coefficients, which are in the order of $10^{-8} – 10^{-6}$ s/m[3, 9] at room-temperature and seem to be much more suitable candidates for devices so far. However, it needs to be mentioned, that the linear ME responses of single-phase materials and composites cannot be easily compared to each other. In order to permit such an analogy the ME coupling-coefficient for composites is usually defined as an AC effect with low AC fields and frequencies between 100 Hz and 1 MHz[1] in contrast to the static effect of single-phase multiferroics.

Certainly the most intensely investigated multiferroic material is $BiFeO_3$ which is ferroelectric ($T_C \approx 1100$ K) and weakly ferromagnetic at room-temperature in thin-film form, whereas for bulk single crystals the weak ferromagnetism is supressed by a cycloidal spin-superstructure.[8, 10] In recent years, tremendous effort has been spent on enhancing its multiferroic properties by modifying it structurally or chemically. It was found, that the FE long-range order is reduced by adding $Bi_{1/2}K_{1/2}TiO_3$ ($T_C \approx 710$ K) to the system and the material, hence, becomes relaxor FE[11] due to quenched cationic charge disorder similarly as, *e.g.*, in the archetypal relaxor $PbMg_{1/3}Nb_{2/3}O_3$[12] and in solid solutions of BKT and LiNbO3,[13] while the magnetic properties are also changed as compared to pure $BiFeO_3$.[14] Here, we show for the first time that relaxor ferroelectrics exhibiting polar nanoregions (PNR) possess the ability to form congruent ferroelectric and magnetic regions, which we will refer to as *multiferroic clusters* (MFC), in the single-phase perovskite $(BiFe_{0.9}Co_{0.1}O_3)_{0.4}$-$(Bi_{1/2}K_{1/2}TiO_3)_{0.6}$ (BFC-BKT). PNR transform into static-PNR (SPNR) below the freezing temperature $T \approx 300°C$ which is characteristic for a non-ergodic relaxor state.[15] The magnetization of the MFC actually arises from ferrimagnetic order, due to coupling of $Fe^{3+}$- and $Co^{3+}$- ions within inherent Fe and Co rich regions. As expected, giant direct and converse ME coupling, $\alpha \approx 1.0 \times 10^{-5}$ s/m (corresponding to a Voltage coefficient $dE/dH \approx 1.3$ kV/(cm Oe)), is observed on MFC on the nano-scale using in-situ PFM under magnetic field and MFM in combination with tip induced poling.

These discoveries open up a new perspective for the class of non-ergodic relaxor multiferroics.

## 2. Results

Polycrystalline, ceramic samples with average grain size of 1.81 μm of the composition $(BiFe_{0.9}Co_{0.1}O_3)_{0.4}$-$(Bi_{1/2}K_{1/2}TiO_3)_{0.6}$ (BFC-BKT) and density of approx. 6.52 g/cm$^3$ were investigated. *X*-ray diffraction revealed that samples are phase pure and have a pseudo-cubic perovskite crystal-structure (see Figure S1). For every multiferroic or magnetoelectric material, both dielectric and magnetic properties are essential. Therefore, these properties will be presented in the following sections before coming to results of ME coupling.

## 2.1. Electrical Characterisation

Polarization *vs.* electric field (*P-E*) loops, permittivity *vs.* temperature as well as PFM images are presented in **Figure 1.**

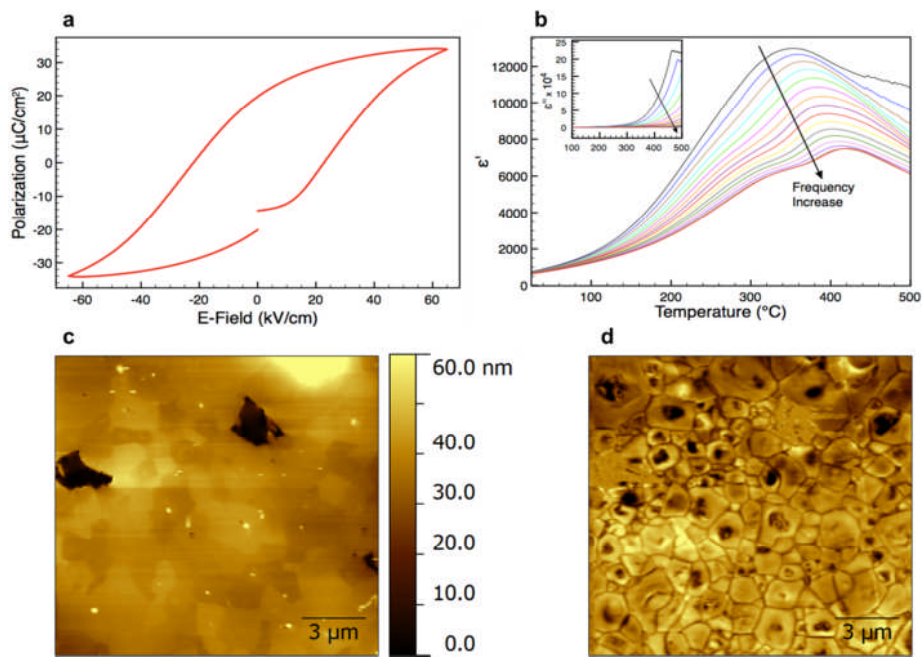

**Figure 1. Electrical characterisation of BFC-BKT ceramics. a**, Polarization *vs.* electric field loop recorded within 1 s at room-temperature. **b**, Permittivity *vs.* temperature curves at logarithmically equidistant frequencies ranging from 1 kHz to 1 MHz. The inset shows an analogous plot of the imaginary part of permittivity. **c**, **d**, Topography and PFM image (X-amplitude) of a polished ceramic. A non-ergodic relaxor state is confirmed by frequency dispersion in **b** and by the presence of static-PNR which occur as dark and bright spots in **d** within a 'matrix' exhibiting lower piezoresponse.

The material displays relaxor ferroelectric behaviour which is evident by the *P-E*-loop in Figure 1a and by the frequency dispersion of the relaxation peak between 350 to 400 °C in permittivity *vs.* temperature curves (Figure 1b). It should be noticed that relaxor-typical PNR have been observed previously in BF-BKT.[11, 16] They are due to random electric fields emerging from the intrinsic cationic charge disorder and are supposed to freeze into a dipolar cluster-glass state on cooling below the low-$f$ peak temperature, $T_m \approx 300°C$.[17]]

On the other hand, reasonably large external electric fields, $|E| > 20$ kV/cm (Figure 1a), suffice to break the glassy disorder and to align all dipolar moments at saturation. This picture is supported by PFM images as shown in Figure 1d, which reveal the presence of SPNR visible as dark and bright spots in Figure 1d[18]. These are usually located in the centre of a grain (compare topography in Figure 1c) and are separated by a dipolar 'matrix' exhibiting low PFM response (corresponding to brown areas), which presumably contains the dynamic PNR. SPNR differ from regular domains, since these are adapted to the crystal lattice via 'ordered' fields and covalent bonds, while SPNR are stabilized by a local excess of a certain component of random electric fields ('random field fluctuation').[19]

**2.2. Magnetic phases**

As mentioned previously, alongside dielectric properties, magnetic properties are key to characterizing multiferroics. This is all the more true for perovskite single-phase multiferroics, whose magnetic properties are usually much weaker than the ferroelectric. Here, the origin of magnetism is often questionable, since small amounts of magnetic secondary phases (~1 wt.%) which are usually not detectable in $X$-ray diffraction, might in many cases account for the observed macroscopic magnetism. Macroscopically, the material displays hysteretic magnetic behaviour at room temperature, which will be presented later (see Figure 4).

Using MFM, magnetic features with sizes ranging from 0.5-1.5 μm were found. **Figure 2**a shows an example of such a feature, which exhibits a magnetic dipolar response, indicating magnetization along an in-plane orientation (see also Figure 6).

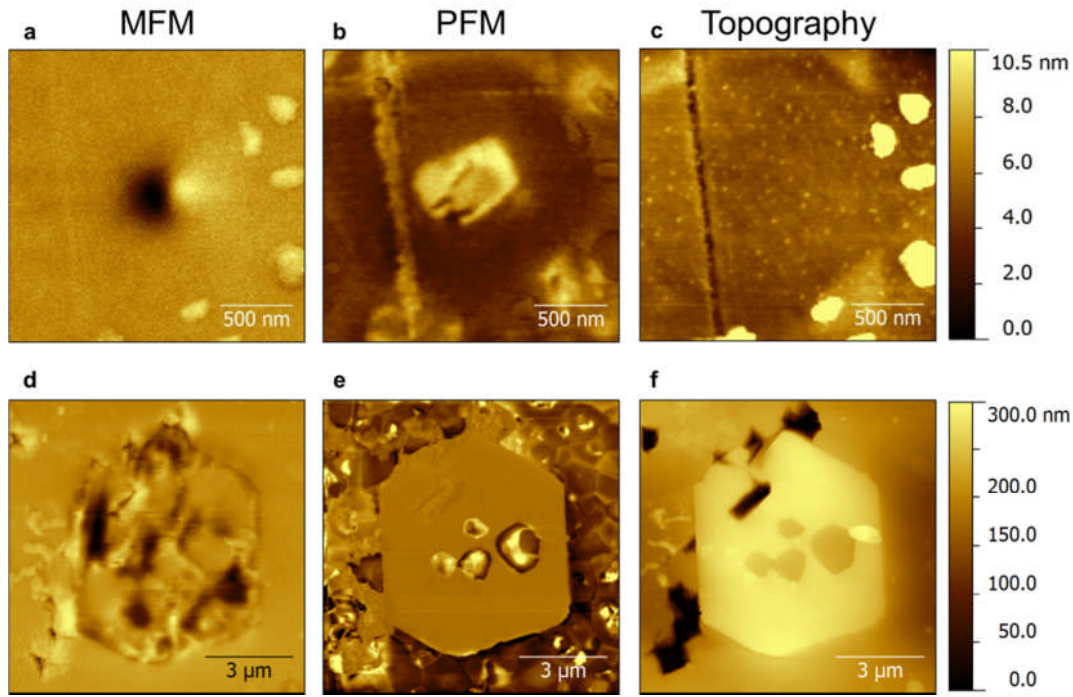

**Figure 2. MFM and PFM images of a multiferroic cluster (top) and of a CoFe$_2$O$_4$ impurity phase particle (bottom). a,b** MFM (phase) and PFM (X-amplitude) images respectively of the very same area showing congruent 'magnetic cluster' and static-PNR with similar shape and size. This multiferroic cluster is not distinguishable from the rest of the material by topography (**c**). A typical CoFe$_2$O$_4$ particle in contrast, is only magnetic and not ferroelectric as evident by MFM (**d**) and PFM (**e**) images respectively, and can be clearly distinguished from the main perovskite phase in topography (**f**).

Most importantly, we found by PFM that for each magnetic feature there was an SPNR in the very same area which had approx. the same size and shape. Due to their congruency, we will refer to them as multiferroic clusters (MFC) henceforth. It will be shown later that the magnetization of the MFC is switchable by magnetic fields, which means that they are in fact magnetic and not due to a measurement artifact (see Figure 6). MFC are both ferroelectric and strongly magnetic based on the strong signals in PFM and MFM respectively. This discovery is very important, since this has not yet been directly observed for a single-phase material to the best of our knowledge.

However, a magnetic impurity phase was also found in the form of micrometre sized particles using MFM (see Figure 2d). Their chemical composition is presumably CoFe$_2$O$_4$ (see

Figure S2) while their volume content is estimated to be below 1% according to SEM-EDX maps. However, MFC are certainly not due to $CoFe_2O_4$ particles since they are multiferroic while $CoFe_2O_4$ particles lack any piezoelectric contrast and thus are clearly not ferroelectric as evidenced by the PFM image in Figure 2e, whereas the surrounding matrix is FE but not FM (see Figure 2d and e).

Furthermore, the $CoFe_2O_4$ particles are considerably larger (approx. 6 μm) than MFC (approx. 1 μm) and can be easily distinguished from the rest of the material by AFM topography since they 'stick out' of the sample surface by several 100 nm (see Figure 3f), apparently due to a lower polishing rate as compared to the rest of the material. In contrast, the MFC cannot be distinguished from the rest of the material by topography (see Figure 3c). Thus, we assume that they have similar mechanical properties and perovskite crystal structure as the vast majority of the material, which is also supported by the fact that they are ferroelectric. Nevertheless, we will refer to the relaxor ferroelectric material that surrounds the MFC as 'matrix', although we expect it to have a very similar crystal structure as the MFC. We will show later, how the chemical composition of MFC and matrix differ (see Figure 4).

But what might be the nature of the MFC? We propose that they are ferrimagnetic due to coupling of $Fe^{3+}$- and $Co^{3+}$- ions which would deliver a net magnetic moment of approx. 1 $\mu_B$ per Fe-Co pair. This agrees with recent neutron diffraction data of Sosnowska et al.[20] on $BiFe_{0.8}Co_{0.2}O_3$ which confirm a ferrimagnetic G-type spin arrangement. Furthermore, we presume the MFC to originate from an increased local concentration of the BFC component which seems to form chemical clusters. In the same area, an SPNR forms as an 'island' of long range FE order within a matrix of disordered PNR, because the BFC component is the one which establishes long range FE order. In contrast, BKT is known to rather induce relaxor properties as in the system $(BiFeO_3)_x$-$(Bi_{1/2}K_{1/2}TiO_3)_{1-x}$.[11]

An idealized crystal structure illustrating ferrimagnetic order of $Fe^{3+}$- and $Co^{3+}$- ions,[20] as well as ferroelectricity due to off-centred B-site ions is shown in **Figure 3**b.

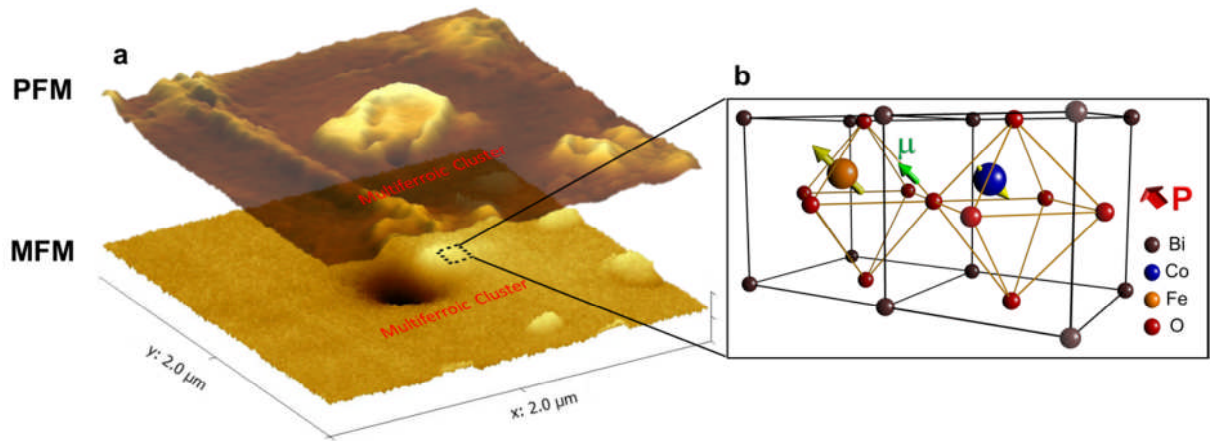

**Figure 3. Proposed structure of a multiferroic cluster (MFC). a**, 3D representation of MFM image (bottom) overlaid with a PFM image of the same sample area showing the MFC. **b**, Idealized crystal structure of the MFC schematically illustrating the ferrimagnetic order of $Fe^{3+}$ and $Co^{3+}$- ions with net magnetization μ and polarization P.

Note that this representation does not display reality truthfully, but is idealized to illustrate the aforementioned facts more clearly.[21] Figure 3a shows the same MFC as in Figure 2 in a 3D representation to illustrate the congruency of the FE and ferrimagnetic regions.

Although a higher local concentration of Fe and Co is expected to be the origin of the MFC, we were not able to resolve it using SEM-EDX analysis, probably because of insufficient sensitivity of the technique (see Figure S3). Therefore, time-of-flight secondary ion mass spectrometry (TOF-SIMS) was employed, which combines an extremely high elemental sensitivity (in the range of ppm and below) with high lateral resolution of down to 50 nm. It is widely used for the spatially resolved determination of dopant concentrations in semiconductors.

Results of SIMS mapping with Bi ion bombardment and positive secondary ions, are presented in **Figure 4**.

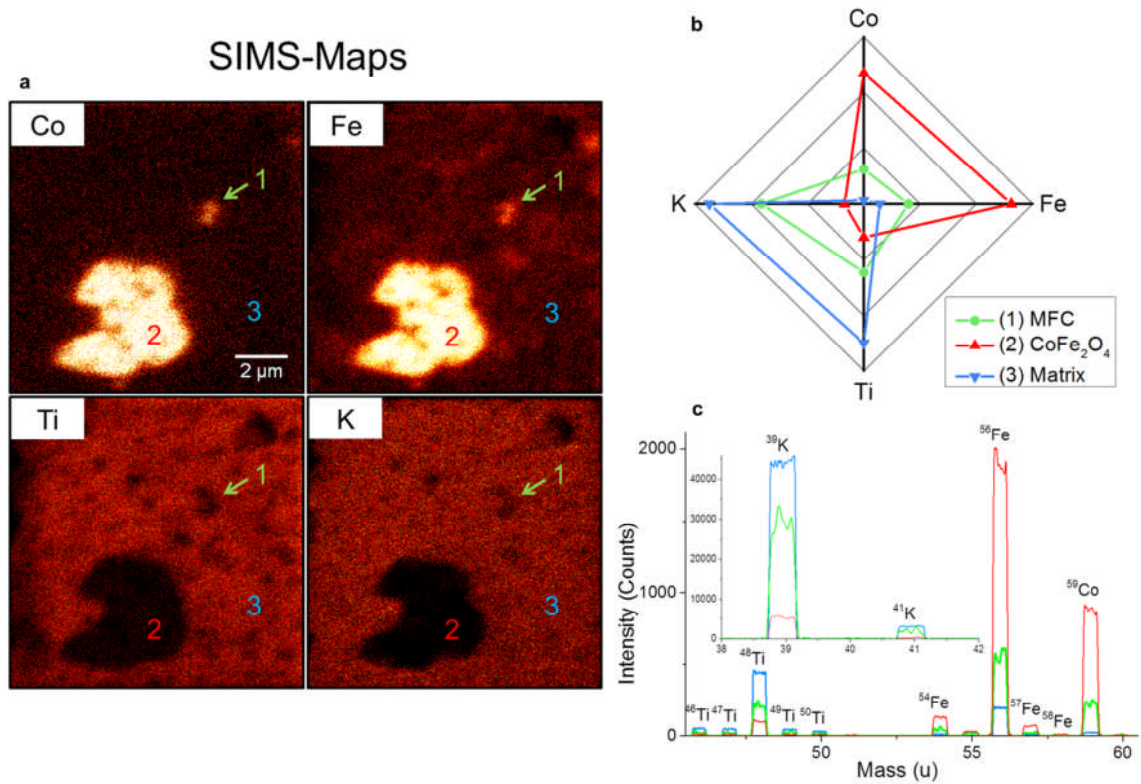

**Figure 4. Secondary ion mass spectrometry (SIMS). a**, Maps of elemental distribution according to labels showing multiferroic cluster (MFC, 1), $CoFe_2O_4$ secondary phase particle (2) and matrix (3). Note that Bi signals could not be analysed since a Bi-ion beam was used as probe. Shape and size of the MFC are in good agreement with MFM and PFM measurements. **b**, Radar chart illustrating relative elemental intensities of the three areas according to mass spectra shown in **c**. As expected, the MFC has an increased Fe and Co content as compared to the matrix but lower K and Ti content. In comparison, the $CoFe_2O_4$ particle has a much higher Fe and Co content than the MFC.

All relevant elements could be identified without problem according to their isotope pattern (see Figure 4c). Maps of elemental composition confirm the existence of Co and Fe rich regions as proposed above.

The maps (Figure 4a) show an area which contains an MFC (marked by green arrow, 1) and a $CoFe_2O_4$ secondary phase particle (2). As expected, the MFC exhibits a higher concentration of Co and Fe, and lower concentration of K and Ti than the surrounding matrix as visible by bright and dark spots in the respective elemental maps. Note that dark spots occurring in all images (*e.g.* top right corner) correspond to pores. The size of approx. 1 µm and the oval shape of the MFC are in good agreement with MFM and PFM measurements. The large feature marked by 2 is identified as a $CoFe_2O_4$ particle by the very high content of Co and Fe as compared to K and Ti and its much larger size of approx. 5 µm typical for these particles (see Figure 2). Note that Bi is not displayed in Figure 4, since a beam of Bi ions was used to analyse the sample surface which resulted in a homogeneous distribution of Bi across the whole area. Using the Bi ion beam was, however, necessary to achieve the necessary resolution. The radar chart in Figure 4b shows relative elemental intensities of Fe, Co, K and Ti for MFC, $CoFe_2O_4$ particle and matrix according to mass spectra shown in Figure 4c (for details of data processing see Supporting Information and Figure S4). In comparison to other areas, the $CoFe_2O_4$ particle almost exclusively contains Fe and Co, whereas the matrix contains mostly K and Ti. In contrast, the MFC contains all elements in medium concentration in agreement with our expectations.

It is important to note, that for $CoFe_2O_4$ the intensities of Fe and Co have approx. the correct ratio of 2:1 (measured 2.23:1) as expected from chemical composition. In SIMS, intensities for different elements usually cannot be directly related to elemental composition due to their different ionization probabilities causing different sensitivity factors for different elements. K as an alkali metal, which is easily ionized, has for example much higher intensities than other elements. However, this is not the case for Fe and Co which have very similar ionization energies (762.5 and 760.4 kJ/mol respectively)[22] and relative sensitivity factors in mass spectrometry.[23] In case of the MFC, the measured ratio of Fe to Co signals is 2.5:1 which is close to the ratio as in $CoFe_2O_4$. Taking into account the lower K and Ti concentration as compared to the matrix, we expect the composition of the MFC to be approx. $(BiFe_{0.7}Co_{0.3}O_3)_{0.6-0.8}$-$(Bi_{1/2}K_{1/2}TiO_3)_{0.4-0.2}$.

SIMS experiments furthermore exclude, that the MFC as imaged by MFM are due to deeper lying magnetic particles, since SIMS is very surface sensitive (1-2 nm penetration depth). After the two different magnetic phases had been characterized on a microscopic level, we also tried to distinguish them on a macroscopic level. To this end, magnetic properties as function of temperature were carefully analysed as illustrated in **Figure 5**.

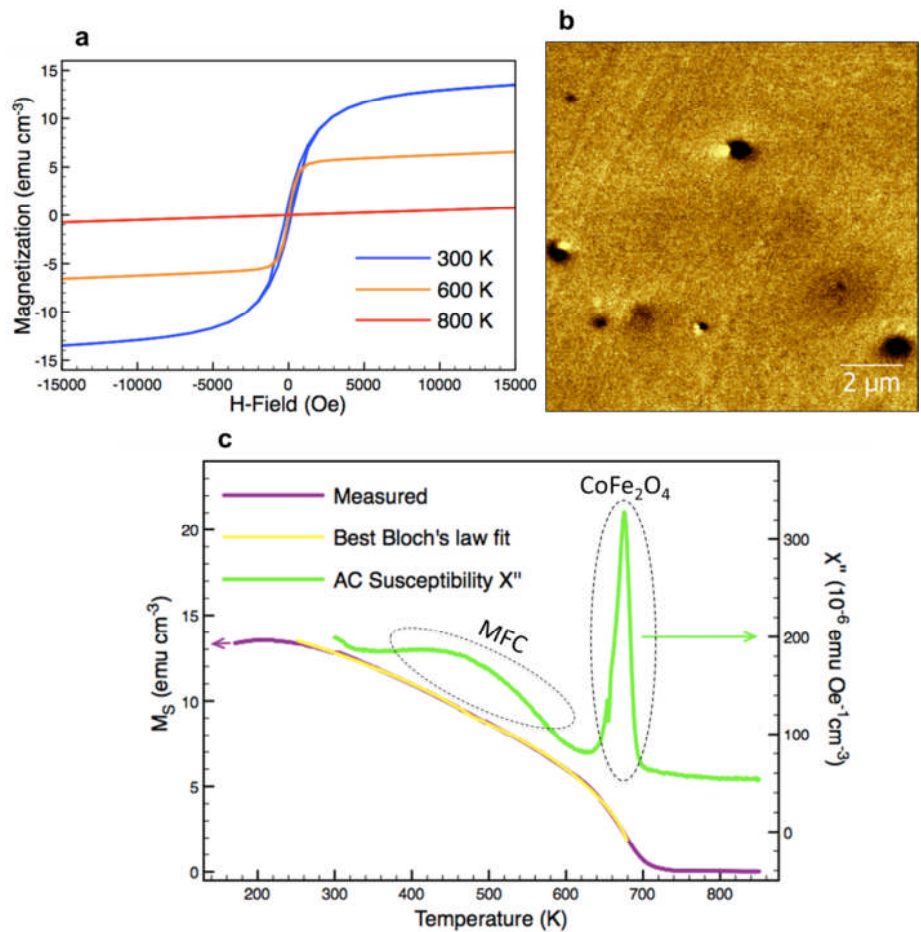

**Figure 5. Magnetic characterisation of BFC-BKT ceramics**. **a**, Magnetization *vs.* magnetic field (*M-H*) loops $T$ = 300, 600 and 800 K. **b**, MFM phase image showing typical MFC. **c**, The spontaneous magnetization, $M_S \approx 2\,M(25\,\text{kOe})\text{-}M(50\,\text{kOe})$ *vs.* $T$ (—) is best-fitted by Equation (1) between 250 and 680 K (—). Imaginary part of AC susceptibility ($\chi''$) *vs.* $T$ (—, $H_{AC}$ = 10 Oe, f = 23 Hz) reveals two peaks at 471.8 and 673.7 K which are correlated with the Curie temperatures of two different magnetic components, MFC and $CoFe_2O_4$.

The magnetization *vs.* magnetic field (*M-H*) loop at room temperature (Figure 5a) displays non-linear magnetic behaviour, with low coercivity and a saturation magnetization of approx. 12.3 emu cm$^{-3}$. The magnetization decreases at higher temperatures until the material becomes paramagnetic (800 K). The spontaneous magnetization, $M_S \approx 2M(25\text{ kOe}) - M(50\text{ kOe})$ *vs. T* (Figure 5c), decreases above approx. 200 K, follows the best-fitted Equation (1) between 250 and 680 K, and shows a Curie transition to a paramagnetic state at $T_C \approx 704$ K (for details of measurement of $M_S$ see Supporting Information). The imaginary part of the AC-susceptibility ($\chi''$) is sensitive to magnetic energy dissipation and thus phase transitions. It exhibits two distinctly different peaks (Figure 5c) and thus indicates two magnetic contributions as expected from previous experiments. The broad peak with center at 471.8 K has a signature that differs significantly from the sharp peak at 673.7 K indicating two magnetic contributions as expected. Hence, we fitted the $M_S$ *vs. T* curve by a function containing two Bloch's Law terms (see Figure 2b):

$$M(T) = M_1(0)\left(1 - \left(\frac{T}{T_{C,1}}\right)^{\frac{3}{2}}\right)^{\beta_1} + M_2(0)\left(1 - \left(\frac{T}{T_{C,2}}\right)^{\frac{3}{2}}\right)^{\beta_2} \quad (1)$$

with $M_i(0)$ = volume-magnetization of contribution *i* at 0 K, $\beta_i$ = critical exponent of contribution *i*

The best-fit parameters are listed in **Table 1:**

| Parameter | Multiferroic clusters | CoFe$_2$O$_4$ |
|---|---|---|
| $M_S(0)$ (emu cm$^{-3}$) | 1.52±0.03 | 14.13±0.03 |
| Critical exponent $\beta$ | 0.786±0.017 | 0.507±0.002 |
| $T_C$ (K) | 478.3±1.39 | 687.7±0.24 |
| Peak temperature of $\chi''$(K) | 471.8±0.30 | 673.7±0.19 |
| Quality of fit | Reduced $\chi^2$ = 0.00175, Adj. $R^2$ = 0.99981 | |

**Table 1. Fitting parameters of Bloch's law fit according to Equation (1).** Curie temperatures of the fits are in good agreement with measured centres of peaks in imaginary AC susceptibility $\chi''$.

The above function fits the $M_S$ vs. $T$ curve very well and Curie temperatures obtained from fitting are in good agreement with maxima of peaks in $\chi''$ (471.8 K vs. 478.3 K and 673.7 K vs. 687.7 K). These temperatures might be interpreted as average Curie temperatures of a given magnetic phase. The relatively small discrepancies in measured and fitted Curie temperatures can be readily explained by different temperature sweep rates used in DC and AC measurements (*i.e.* sample's temperature lagging the sensor) and/or the field dependence of $T_C$ ($M_S$ measured at 25 and 50 kOe while $\chi''$ at 10 Oe).

We attribute the contribution with $T_C$ around 680 K to the $CoFe_2O_4$ secondary phase which, as a classical magnetic material, is expected to exhibit a sharp peak in $\chi''$ at the Curie transition. Although the Curie temperature of pure $CoFe_2O_4$, 793 K,[24] is higher, this might be explained by doping with diluent diamagnetic ions such as $Ti^{4+}$, as measured by SIMS (see Figure 4). The broad peak around 475 K on the other hand, is attributed to the MFC. A lower $T_C$ for the MFC as compared to pure $BiFeO_3$ with its Néel temperature of 650 K due to doping with diamagnetic $K^+$ and $Ti^{4+}$ ions is also expected in this case. Since the MFC presumably have a perovskite crystal structure which is diluted with diamagnetic ions especially at the edges (see Figure 4), they should at least partly consist of a network of magnetic ions, following percolation statistics similar as in the well-studied $PbFe_{1/2}Nb_{1/2}O_3$ (PFN).[25] Such clusters are expected to exhibit a broad distribution of Curie-temperatures as evident from the broad peak in $\chi''$ vs. $T$ due to statistics. This is also reflected in the large critical exponent leading to a relatively linear $M_S$ vs. $T$ curve, whereas $CoFe_2O_4$ exhibits a rather classical Curie transition with a critical exponent close to 0.5.

Although, the macroscopic $M_S$ of the sample is dominated by the CoFe$_2$O$_4$ phase, the contribution of the MFC (approx. 10%) cannot be neglected, since the $M_S$ vs. $T$ curve is fitted much less accurately by just one Bloch's law term. However, it will be shown in the following paragraphs, that the MFC show strong local ME coupling, which is not influenced by the secondary phase.

### 2.3. Converse Magnetoelectric Coupling

After elucidation of the material's magnetic properties, we started to investigate ME coupling in the multiferroic cluster previously presented in Figure 3a. Results are shown in **Figure 6**.

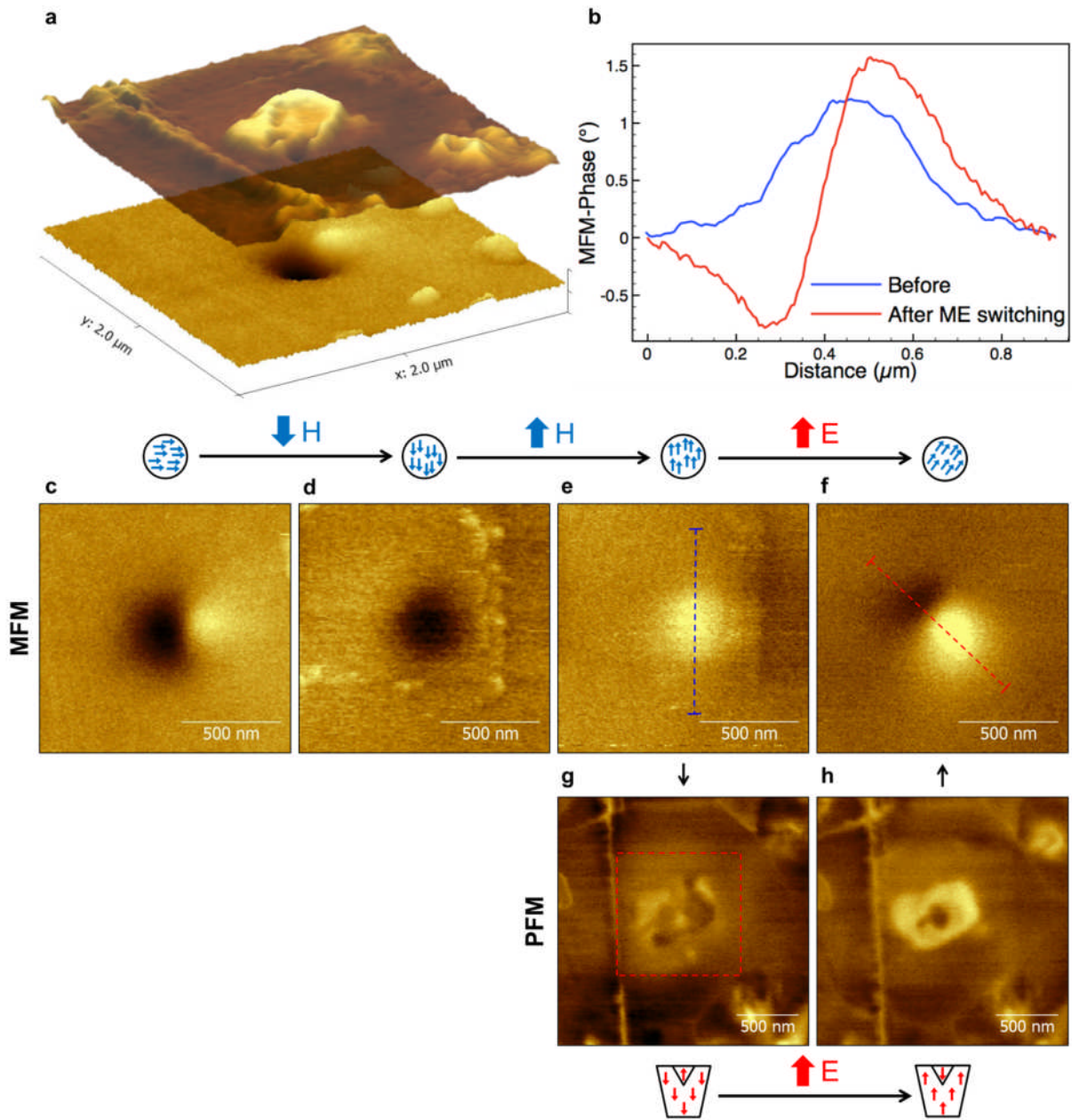

**Figure 6. MFM measurements in combination with electric field poling. a**, 3D representation of the same MFC as in Figure 2 and 3. **c,d,e**, MFM images of the MFC as in **a**, before (**c**) and after ex-situ switching with out-of-plane magnetic fields of ±0.48 T (**d,e**) as indicated by blue arrows. Symbols above images, represent single magnetic domains according to the dipolar magnetic MFM response. **g,h**, PFM images of MFC recorded after ex-situ magnetic switching, before (**g**) and after (**h**) electric poling by scanning a rectangular area as indicated by red dashed rectangle while applying a DC bias. Configurations of MFC's polarization are illustrated by symbols below PFM images. **f**, MFM image after electric field poling, showing ME switching (compare **e**). **b**, MFM phase cross-sections of MFC before and after ME switching across the blue and red dashed lines in **e** and **f** respectively.

As mentioned earlier, the MFC exhibits a magnetic dipolar response as for a single domain particle, with an in-plane orientation of the magnetization as illustrated above the image.[26] Using ex-situ magnetic fields which were applied outside the sample environment of the atomic force-microscope, we switched the MFC (Figure 6d and e), and thus proved that it is in fact magnetic and not an imaging artifact. More importantly, we were able to switch the MFC by poling it with an electric field. Figure 6g and h show the dielectric structure of the MFC before and after poling, by scanning a rectangular area around it as indicated by the red dashed rectangle while applying a DC voltage of 20 V. For imaging an AC voltage of 2 V was applied. After electric poling, the MFC's magnetization was switched from a complete out-of-plane (Figure 6e) to a partly in-plane (Figure 6f) orientation as indicated by symbols. This experiment could be reproduced very accurately. For a discussion of possible artifacts, see Supporting Information.

### 2.4. Direct Magnetoelectric Coupling

Additionally to the converse ME effect, we also investigated the direct effect on the micro-scale using in-situ PFM under magnetic field which is illustrated in **Figure 7**.

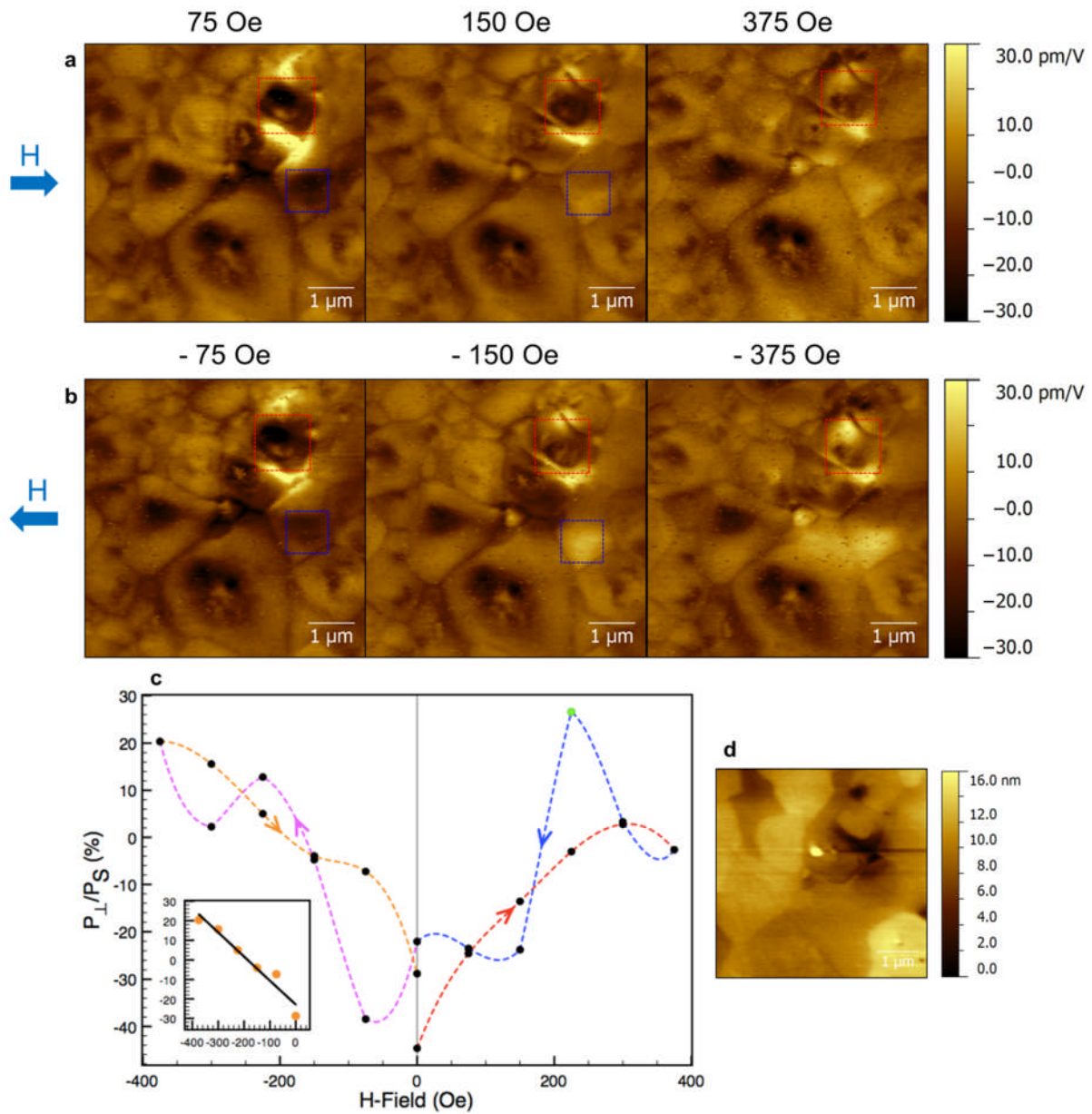

**Figure 7. In-situ PFM under magnetic field experiments. a,b** Selected out-of-plane PFM images from a magnetic-field loop series consisting of 21 images. Magnetic field strength is according to labels and direction is according to blue arrows. Change of piezoactivity is evident *e.g.* in regions marked by blue and red rectangles. **c**, Polarization *vs.* magnetic field data referring to the region marked by the red rectangle in PFM images. Dashed lines are spline-fits of data points (•) to indicate the direction of the magnetic field. A strong sporadic switching event is marked by (•). Inset shows linear fit to data points from the orange line. **d**, AFM topography corresponding to **b** row images.

Figure 7 shows selected out-of-plane in-situ PFM images under magnetic field. In total 21

PFM images constituting a complete magnetic field loop with 0, ±375 and 75 Oe as starting-point, magnetic field range and step-size respectively, were recorded. Images in row **a** clearly display a magnetic field-induced change of an MFC, in the top right corners of the images. The direction of magnetic field, relative to the sample surface is indicated by blue arrows beside the PFM images. Other regions exhibiting strong ME switching are marked by blue rectangles.

By taking the average PFM-signal of a highly ME active region marked by red rectangles in row **b**, the change of out-of-plane polarization of this area as a function of magnetic field can be obtained which is shown in Figure 7c (for calculation of $P_\perp/P_S$ from PFM signals see below). This plot reveals a large change of polarization over a relatively small magnetic field range. The curve exhibits a V-like shape, where the polarization varies roughly linearly with magnetic field but, regardless of its direction.

If we consider that the underlying ME coupling mechanism is linear and intrinsic, the V-shape of the curve might be explained by the fact that the MFC exhibits a low magnetic coercivity. Thus, the linear ME coupling coefficient might change its sign at low fields when inverting the magnetic field direction.

The curve shape might also be explained by a stress-strain-mediated ME coupling mechanism, via magnetostriction and piezoelectricity. Many magnetostrictive materials such as ferrite spinels or metal alloys exhibit such a V-shape magnetostriction curve.[3] The resultant magnetostrictive strain in turn would be coupled to polarization linearly via the piezoelectric effect and thus would explain the observed shape of the curve.

Furthermore, strong sporadic switching events were observed (see green dot in Figure 7c). This behaviour was also reported by Evans *et al.*,[27] who observed domain-switching by PFM with ex-situ magnetic fields in multiferroic $(PbZr_{0.53}Ti_{0.47}O_3)_{0.6}$–$(PbFe_{0.5}Ta_{0.5}O_3)_{0.4}$. The authors attributed this behaviour to sudden releases of elastic energy.

Since the ME response appears to be roughly linear within the experiment's uncertainty in a small range of magnetic fields and for the sake of quantifying the effect in a comparable way, we will estimate the linear ME coupling-coefficient for a limited range of magnetic fields, although the coupling is clearly non-linear over the whole range of magnetic fields tested in this experiment. The direct linear ME coupling-coefficient can be expressed as[2]

$$\alpha_{ij} = \frac{\partial P_i}{\partial H_j} \qquad (2)$$

with $P_i$ = electrical polarization and $H_j$ = magnetic field components, respectively.

To estimate the change of polarization over a range of magnetic fields, PFM signals need to be quantified. A calibration factor $K_{BFC}$ connecting PFM signals to a polarization change for BFC-BKT can be obtained by using a standard calibration sample, periodically poled lithium niobate (PPLN) together with macroscopic piezoelectric $d_{33}$ coefficients (for details and discussion of the calibration see Supplementary Information).

$K_{BFC}$ can be used to estimate the ME coupling-coefficient as follows:

$$\alpha = \frac{2 P_s \cdot m}{K_{BFC}} \qquad (3)$$

where $P_s$ is 33.9 µC/cm$^2$ and is obtained from a macroscopic *P-E*-loop (see Fig. 1a) measurement and $m/K_{BFC}$ = 0.123 %/Oe is the slope of the linear fit to the orange data points in Figure 7c (inset) and corresponds to the change of polarization with magnetic field. Thus we estimate an effective coefficient $\alpha_{31}^{eff}$ = 1.0 x 10$^{-5}$ s/m which is, to the best of our knowledge, the highest coupling coefficient reported for a single-phase multiferroic yet. It is roughly two orders of magnitude larger, than the local coupling coefficient estimated by Evans et al.[27] and five orders larger than the macroscopic effect obtained on (BiFeO$_3$)$_{0.6}$–(Na$_{0.5}$Bi$_{0.5}$TiO$_3$)$_{0.4}$.[28]

Nevertheless, we want to stress that the calculated value is an estimate of the order of magnitude rather than an exact determination of the ME coefficient. It is, however, intuitive that the coefficient should be large, since relatively small magnetic fields (375 Oe at most) result in considerable switching of the MFC, which should be due to an extremely large ME coupling coefficient.

The strong ME coupling correlates well with the fact that MFC exhibit both ferroelectric and presumably ferrimagnetic order. Another reason for the exceptionally large ME coupling, might be the dielectrically flexible matrix, surrounding the MFC. The dynamic and flexible PNR might facilitate ME reorientation of the MFC's polarization by accommodating strain due to the reorientation process, which effectively reduces clamping of the MFC. In case of a large scale single domain multiferroic material, this might not be possible.

We will also try to compare the direct to the converse ME coefficient from the experiment presented in Figure 6. The converse linear ME coupling-coefficient $\alpha_c$ can be expressed as following:

$$\alpha_{c,ij} = \mu_0 \frac{\partial M_i}{\partial E_j} \quad (4)$$

with $\mu_0$ = vacuum-permeability, $M_i$ = volume-magnetization, and $E_j$ = electric field components.

Based on the magnitude of MFM signals, we can assume that the magnetization is reasonably large. Numerical values for $\mu_0 M$ and $P$ for the MFC from Figure 6 and Figure 7 (marked by red rectangle), respectively, are expected to be similar when converted to SI-units although $P$ will be surely larger. Unfortunately, the electric field used for poling in Figure 6i is unknown due to the non-uniform field underneath the tip. However, we can expect a large ME coupling coefficient due to the large reorientation of approx. 46.8% of the MFC's magnetic moment upon application of electric field. The reorientation can be estimated from the MFM cross section through the MFC (Figure 6b) by comparing the relative heights of signals corresponding to bright and dark areas. This reorientation is almost identical as compared to that of the polarization displayed in Figure 6c over a change of magnetic field of 375 Oe. Therefore we expect a similar order of magnitude for the direct and converse ME coupling coefficients.

## 3. Conclusions

We could show that inherent multiferroic clusters (MFC) exist in BFC-BKT which are ferroelectric and presumably ferrimagnetic due to an increased concentration of Fe and Co as observed by PFM, MFM and SIMS. While ferrimagnetism stems from antiparallel alignment of Fe and Co atomic moments, the non-ergodic relaxor ferroelectric properties of the material enable long range FE order in the very same region due to a lower concentration of the charge disorder inducing component BKT. Thus formation of MFC is governed by the ferrimagnetic BFC component and the relaxor-state inducing BKT component which presumably also improves the dielectric properties of the solid solution. However, the non-ergodic relaxor nature does not only enable formation of MFC, but is also expected to be partly responsible for the exceptionally large ME coupling. Dynamic PNR surrounding the MFC form a dielectrically and mechanically flexible environment, which presumably facilitates reorientation of the MFC's polarisation by accommodating strain. Thus, clamping of MFC is effectively reduced.

A caveat is the formation of small amounts of $CoFe_2O_4$ secondary phase which should be eliminated in future work. We were able to distinguish magnetic contributions due to MFC and $CoFe_2O_4$ by carefully analysing DC and AC magnetometry data, as a function of temperature. This approach should be also applicable to other single-phase multiferroics where magnetic secondary phases usually pose a problem for magnetic characterisation. We can exclude without doubt that $CoFe_2O_4$ particles were erroneously regarded as MFC, nor did $CoFe_2O_4$ particles exhibit the observed local ME coupling effects.

We expect our findings to spark significant new research in this new class of non-ergodic relaxor multiferroics, also on thin film, single crystal or oriented ceramic materials, especially since the material is lead free and consists only of abundant elements. Compositions closer to the one estimated for the MFC, $(BiFe_{0.7}Co_{0.3}O_3)_{0.6-0.8}$-$(Bi_{1/2}K_{1/2}TiO_3)_{0.4-0.2}$, might be ideal starting points for such experiments. Since the ME coupling is restricted to well-separated magnetic regions, applications of electrically addressable magnetic MFC might be envisaged for future ME random access memory devices (MERAM)[5] based on BFC-BKT thin films. They are suitable for fulfilling the dream of an electrically controlled magnetic nanodot storage device.[29] Such a device is schematically illustrated in **Figure 8**.

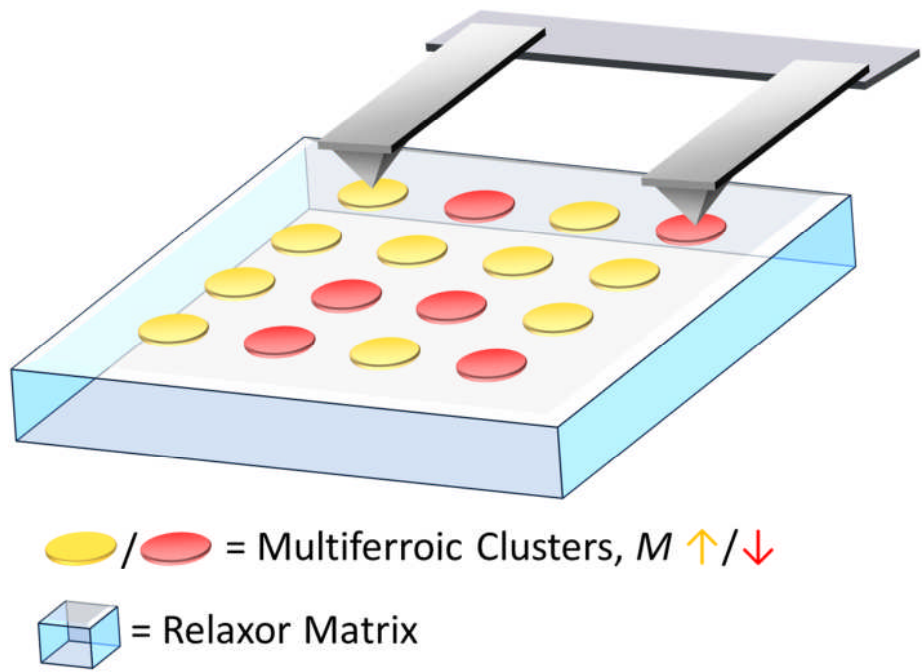

**Figure 8. Schematic illustration of a possible ME memory device.** An epitaxial BFC-BKT thin-film might be engineered to contain an array of multiferroic clusters within a relaxor ferroelectric matrix. Information is stored in an electrically controlled magnetic bit, adressable e.g. via cantilever-type probes in a microelectromechanical systems (MEMS) device.

## 4. Experimental Details

BFC-BKT ceramic pellets were prepared analogously to [30]. As starting materials $Bi_2O_3$, $Fe_2O_3$, $TiO_2$, CoO (all Sigma Aldrich) and $K_2CO_3$ (Alfa-Aesar) (all 99.9% purity) were mixed in appropriate proportions and a Dynomill Typ KDLA by Willy A Bachofen was used for milling. Pellets were sintered at 1065 °C for 2 hours.

For electrical testing, silver electrodes were applied onto coarsely polished samples at 550 °C using silver paint. *P-E*-loops and permittivity *vs.* temperature measurements were carried out on a Radiant Precision 10kV HVI II and a HP 4284 A Precision LCR Meter in junction with a tube furnace respectively.

For magnetic measurements and *X*-ray diffraction, the powder of sintered and crushed pellets was used and measurements were carried out on a SQUID-VSM (MPMS 3) by Quantum Design using the VSM oven option and a Phillips X'PERT respectively.

For scanning probe-microscopy experiments, ceramic samples were polished using a Motopol 2000 by Buehler and Buehler polishing products. A surface roughness of around 1 nm was achieved by polishing in several steps using various polishing-cloths in combination with diamond abrasive-liquids where the diamond particle-size was gradually reduced for consecutive steps, until a final polishing step involving 200 nm sized colloidal silica particles was reached.

PFM and MFM experiments were carried out on a 5420 AFM by Agilent Technologies with the MAC Mode III extension.

For PFM experiments, DCP11 conductive-diamond coated tips by NT-MDT were used. The tip was electrically grounded whilst a 'bottom-electrode' underneath the sample and in electrical contact with it, was biased. All PFM imaging was carried out at a frequency of 70 kHz of the AC voltage.

To avoid distortions by an inherent background-signal,[31] X-amplitude (often referred to as mixed signal) was recorded instead of R-amplitude (often referred to as amplitude) and Phase, while Y-amplitude was minimized by applying a phase-shift between reference and measured signal electronically.

Magnetic fields for in-situ under magnetic field PFM experiments were generated by the Magnetic Lateral Field Module 5420 by ScienTec with magnetic fields of up to ±750 Oe. PPP-MFMR AFM tips by Nanosensors were used for MFM experiments in a constant frequency mode. Additional MFM measurements were carried out on an AttoMFM I by Attocube. A TOF.SIMS 5 by ION-TOF was used for SIMS measurements.

**Supporting Information**

Supporting Information is available from the Wiley Online Library or from the author.

**Acknowledgements**

L. F. H acknowledges project funding by the European Commission through the ITN NANOMOTION (PITN-GA-2011-290158).

Furthermore, he wishes to express his gratitude to E. Soergel and V. Shvartsman for many fruitful discussions and to M. Fenner from Keysight Technologies (formerly Agilent Technologies) for technical support.

L. F. H. thanks Philippa M. Shepley for Magneto-optical Kerr measurements.

Processing of AFM-images was done using the free AFM software 'Gwyddion'[32] for which excellent user support by D. Nečas is gratefully acknowledged.

O.C. acknowledges grant EP/K00512X/1 which enabled SQUID-VSM measurements.

D. C. L. thanks the Deutsche Forschungsgemeinschaft (DFG) for partial support through Forschergruppe 1509, "Ferroische Funktionsmaterialien" (Lu729/12).

H. W., J.L. and S.S. thank the Deutsche Forschungsgemeinschaft (DFG) for partial support through Schwerpunktprogramm 1681, „Feldgesteuerte Partikel-Matrix-Wechselwirkungen" (WE2623/7-1) and Stiftung Mercator (MERCUR).